# Numerical simulation of breathers interactions in two-dimensional O(3) nonlinear sigma model

F. Sh. Shokirov

*S. U. Umarov Physical-Technical Institute of Academy of Sciences  
of the Republic of Tajikistan, Aini Avenue 299/1, Dushanbe*

**Abstract.** By numerical simulation methods the interactions of oscillating solutions (breathers) of the (2+1)-dimensional O(3) nonlinear sigma model is investigated. The models of head-on collisions in which the interacting breathers, in particular, are combined into a single oscillating soliton, the reflections from each other and pass through each other are obtained.

## I. Introduction

This paper presents the results of a series numerical simulations of a head-on collisions of breather solutions in the (2+1)-dimensional anisotropic O(3) nonlinear sigma model (NSM).

Recall that the Euler-Lagrange equations of O(3) NSM are as follows [1-7]:

$$\frac{\partial^2 \theta}{\partial t^2} - \frac{\partial^2 \theta}{\partial x^2} - \frac{\partial^2 \theta}{\partial y^2} + \left[\delta - \left(\frac{\partial \varphi}{\partial t}\right)^2 + \left(\frac{\partial \varphi}{\partial x}\right)^2 + \left(\frac{\partial \varphi}{\partial y}\right)^2\right] \sin\theta \cos\theta = 0, \quad (1)$$

$$2\cos\theta \left(\frac{\partial \theta}{\partial t}\frac{\partial \varphi}{\partial t} - \frac{\partial \theta}{\partial x}\frac{\partial \varphi}{\partial x} - \frac{\partial \theta}{\partial y}\frac{\partial \varphi}{\partial y}\right) + \sin\theta \left(\frac{\partial^2 \varphi}{\partial t^2} - \frac{\partial^2 \varphi}{\partial x^2} - \frac{\partial^2 \varphi}{\partial y^2}\right) = 0,$$

where $\theta(x,y,t)$ and $\varphi(x,y,t)$ – are Euler angles, associated with isospin parameters of model (1) as follows:

$$s_1 = \sin\theta \cos\varphi, \qquad s_2 = \sin\theta \sin\varphi, \qquad s_3 = \cos\theta,$$

$$s_i s_i = 1, \qquad i = 1,2,3;$$

$\delta$ – are anisotropy constant. Fields $s_1$, $s_2$, $s_3$ in this case are the coordinates of the unit isovector $S(s_1, s_2, s_3)$, which describes the dynamics of the solutions in isotopic space of the sphere $S^2$.

Note also those equations (1) in the meridian intersection of the sphere $S^2$ ($\varphi(x,y,t) = 0$) are reduced [3-7] to the (2+1)-dimensional sine-Gordon equation (SG) in the following form:

$$2\left(\frac{\partial^2 \theta}{\partial t^2} - \frac{\partial^2 \theta}{\partial x^2} - \frac{\partial^2 \theta}{\partial y^2}\right) = -\delta \sin 2\theta. \quad (2)$$

In [7] (preprint) the analytical form of test functions of the equation (2), which evolve to stable in time periodic radially-symmetric solutions has been established. In the terms of isospin fields the obtained solutions have the following form:

$$s_1 = -\frac{2\xi}{1+\xi^2}\cos\varphi, \qquad s_2 = -\frac{2\xi}{1+\xi^2}\sin\varphi, \qquad s_3 = \frac{1-\xi^2}{1+\xi^2}, \quad (3)$$

$$\xi(x,y,t) = \frac{\lambda}{\sqrt{1-\lambda^2}} \frac{\sin\varphi}{\cosh(\lambda x)\cosh(\lambda y)},$$

where $\lambda(t)$ and $\varphi(t)$ determined by the equations

$$\lambda_{tt} + \frac{3\lambda}{2(1-\lambda^2)}\lambda_t^2 - \lambda(2-\lambda^2)\varphi_t^2 - \frac{12\sqrt{1-\lambda^2}}{\lambda^2}\left((1-\lambda^2)\text{arcth}(\lambda) - \lambda\right) = 0,$$

$$\varphi_{tt} + \frac{2-\lambda^2}{\lambda(1-\lambda^2)}\lambda_t\varphi_t = 0,$$

obtained in [7]. In this paper, these solutions were named breather (Fig.1).

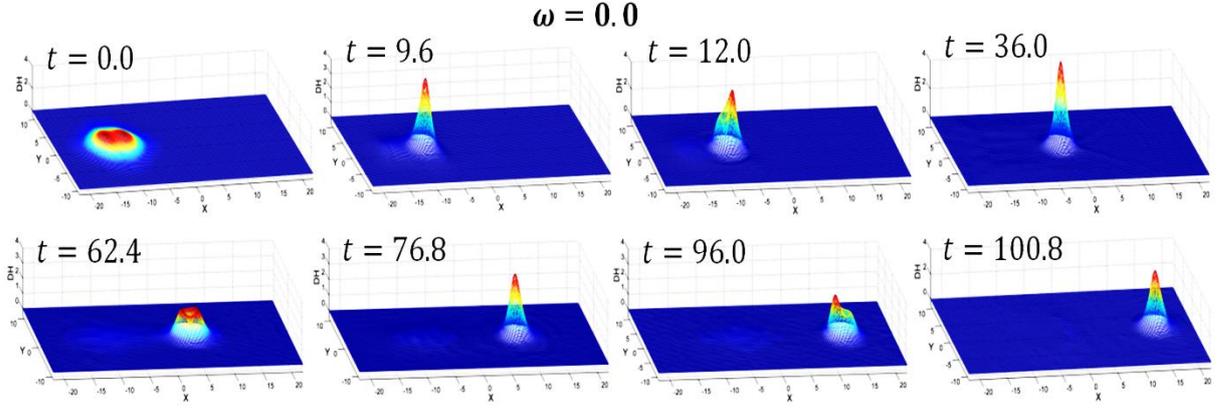

**Fig. 1.** Evolution of energy density (DH) moving at speed $v_{t_0} \approx 0.707$ ($t_0 = 0.0$) breather (3) of the SG equation (2) ($\omega = 0.0$, $\varphi(t) = 0.0$). Simulation time: $t \in [0.0, 100.8]$.

On the basis of the solutions found by adding the perturbations ($\omega \neq 0.0$, $\varphi(t) \neq 0.0$) to vector of A3-field in isotopic space of $S^2$, have been obtained solutions for the O(3) NSM (Fig.2).

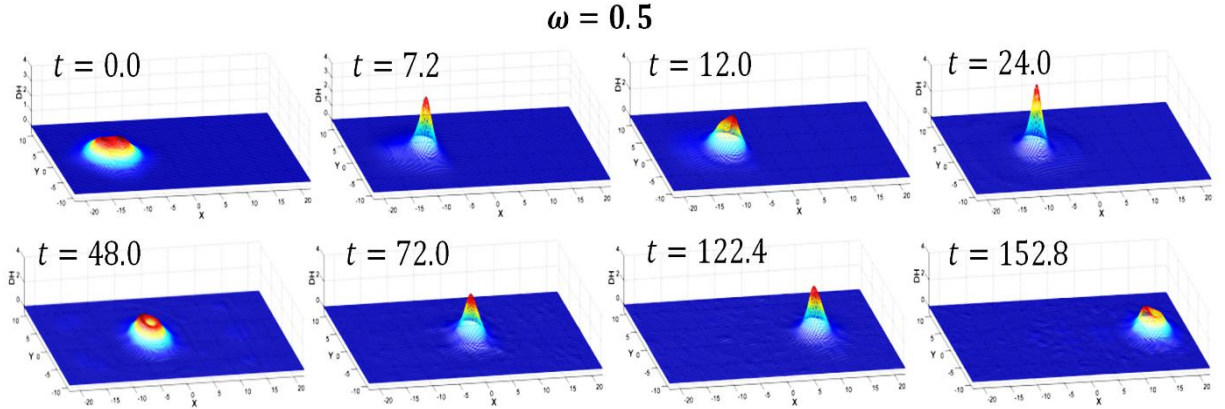

**Fig. 2.** Evolution of energy density (DH) moving at speed $v_{t_0} \approx 0.707$ ($t_0 = 0.0$) breather (3) of the O(3) NSM (1) ($\omega = 0.5$, $\varphi(t) = -0.5\tau$). Simulation time: $t \in [0.0, 152.8]$.

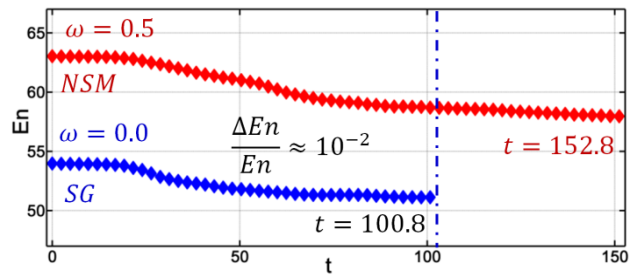

**Fig. 3.** The values of the energy integral (En) of moving at speed $v_{t_0} \approx 0.707$ ($t_0 = 0.0$) breathers (3) of the O(3) NSM (1) at $\omega = 0.0$ (♦) and $\omega = 0.5$ (♦). The total simulation time: $t \in [0.0, 152.8]$.

The energy integral of obtained models is preserved with accuracy: $\frac{\Delta En}{En} \approx 10^{-3} - 10^{-2}$ (Fig.3).

Some changes of breather solutions speed in this case was discussed in [7]. Numerical experiments have shown that speed of breathers motion $v_t$ ($t > 0.0$) always less than their initial speed $v_t < v_{t_0}$ ($t_0 = 0.0$). Obviously, in the case of breathers, some of their energy in the early evolution ($En(v_{t_0})$) absorbed by the characteristic oscillatory dynamics. On Fig.1 breather of the SG equation (2) at $t = 100.8$ passes a distance equal to $s \approx 26.5$ units (average speed: $v_t \approx 0.263$) and loss of speed is $v_{loss} \approx 0.628\%$. This same distance ($s \approx 26.5$) breather of the O(3) NSM passes at $t = 152.8$ (Fig.2) and thus, the overall loss of speed in this case is $v_{loss} \approx 0.75\%$. Note that in the case of breathers of the O(3) NSM to attenuation of their speed besides the oscillating motion affect also the perturbations ($\varphi(x, y, t) = \omega\tau$, $\omega \neq 0.0$) an added to the dynamic of isotopic spin in the space of sphere $S^2$.

At the evolution the test solutions (3) by reset the excess energy in the form of linear perturbation waves acquire relatively stable dynamics (Fig.3). Note that in all numerical experiments of the present paper, has been used the special boundary conditions, which absorbing the linear wave disturbances, emitted by breather solitons [1-7].

Numerical experiments of this paper are based on algorithms and difference schemes [8], which developed in works [1,2,6] for the stationary case, where have been used the properties of the stereographic projection, taking into account the features of theoretical-groups constructions of O(N) NSM class of the field theory (see., e.g., [1-7]).

In the second and third part of the paper presents the results of research into the processes of interactions of the single type breathers (3) of SG equation (2) and O(3) NSM (1) respectively. The models of breather solutions association into a single oscillating structure, as well as models where interacting breathers pass through each other are obtained. In the fourth part of the paper shows the results of numerical simulation of interaction breathers of the O(3) NSM, which differ by mutually inverse values of the frequency of A3-field vector rotation ($\omega_1 = -\omega_2$) in isospace of the sphere $S^2$ [3-7]. In this case, has been obtained a collisions model and reflections of breathers, as well as a model where there is a gradual destruction of one of the breathers. The results of the interaction of breathers (3) of the O(3) NSM (1) and SG equation (2) are given in the fifth part of present paper. The models, where occurs the process of unification of interacting solitons by through "absorption" and the destruction the breather of the SG equation by breather field of the O(3) NSM are obtained. In the last section of the paper discusses the properties of the obtained results and their possible application to the study of other tasks.

Theoretical calculations of processes of formation and evolution of the two-dimensional breathers were carried out in [7]. This paper presents statistical data obtained as a result of numerical simulations of the interaction processes of test solutions [3], at different parameters.

**II. Interaction of breathers in the sine-Gordon equation**

In this part, as a test problems were obtained model of the head-on collisions of two-dimensional breathers (3) of the SG equation (2) at different initial values speeds of their motion:

1. $\vec{v}_1(t_0) \approx 0.287 \rightarrow\leftarrow \tilde{v}_2(t_0) \approx 0.287$;
2. $\vec{v}_1(t_0) \approx 0.447 \rightarrow\leftarrow \tilde{v}_2(t_0) \approx 0.447$;
3. $\vec{v}_1(t_0) \approx 0.707 \rightarrow\leftarrow \tilde{v}_2(t_0) \approx 0.707$;

which given by the Lorentz transformations (at $t_0 = 0.0$).

**2.1.** $v_{12}(t_0) \approx 0.287$ (Fig.4). In a collision ($80 < t < 95$) breathers are combined into a single oscillating localized structure. Next, at $t = 99.6$ occurs a short-term separation of breathers, at this observed the radiation a certain portion of their energy. In between $t \in (120.0, 140.0)$ breathers again evolve as a single oscillating soliton. At the $t \in (140.0, 160.0)$ we again observed the separation of breathers in the transverse direction and the radiation of a certain part of their energy. When $t \in (160.0, 180.0)$ again observed association oscillating solitons. Next, when $t \in (190.0, 210.0)$ occurs a temporary separation and unification of breathers, which accompanied by emission of a certain part of their energy. When $t > 210.0$ the breathers again separated and emit a certain part of energy.

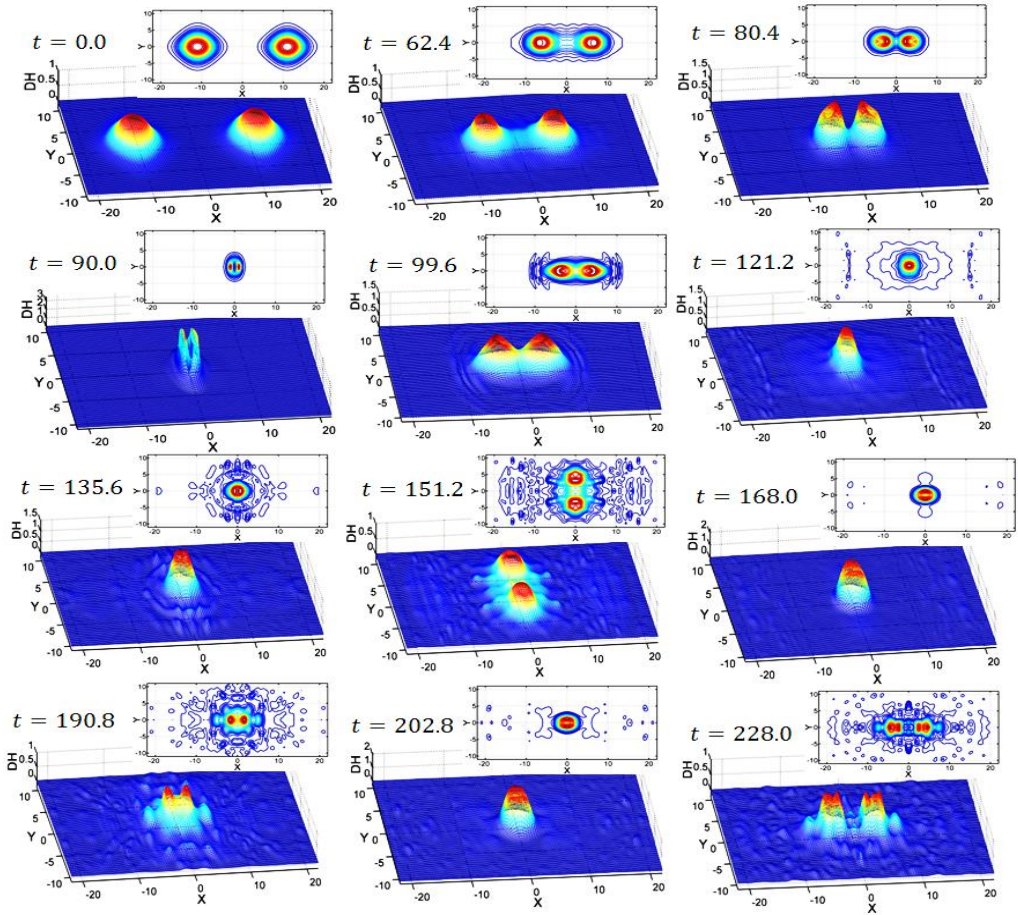

**Fig.4.** Head-on collisions of breathers (3) of the SG equation (2), moving at speed $v_{12}(t_0) \approx 0.287$, ($t_0 = 0.0$). Evolution of energy density (DH). Simulation time: $t \in [0.0, 228.0]$.

The energy integral of system of interacting breathers preserved with accuracy: $\frac{\Delta En}{En} \approx 10^{-2}$ (Fig.7).

Thus, in this experiment, at the frontal collisions of breathers of the SG equations observed their unification into a localized single structure with periodic dynamics.

The following subparts describe similar experiments where the initial speed of motion of the interacting breathers has been increased.

**2.2.** $v_{12}(t_0) \approx 0.447$ (Fig.5). In this case, both breather move in the opposite direction at a speed $v_{12}(t_0) \approx 0.447$, given at $t = 0.0$. When $t \approx 50.0$ breathers passes distance equal to $S_{12} \approx 7.0$ units, collide and form a single oscillating soliton. Formed oscillating soliton during the interval $t \in (55.0, 72.0)$ periodically emits a significant part of its energy and since the time of $t \approx 75.0$ gradually destroyed.

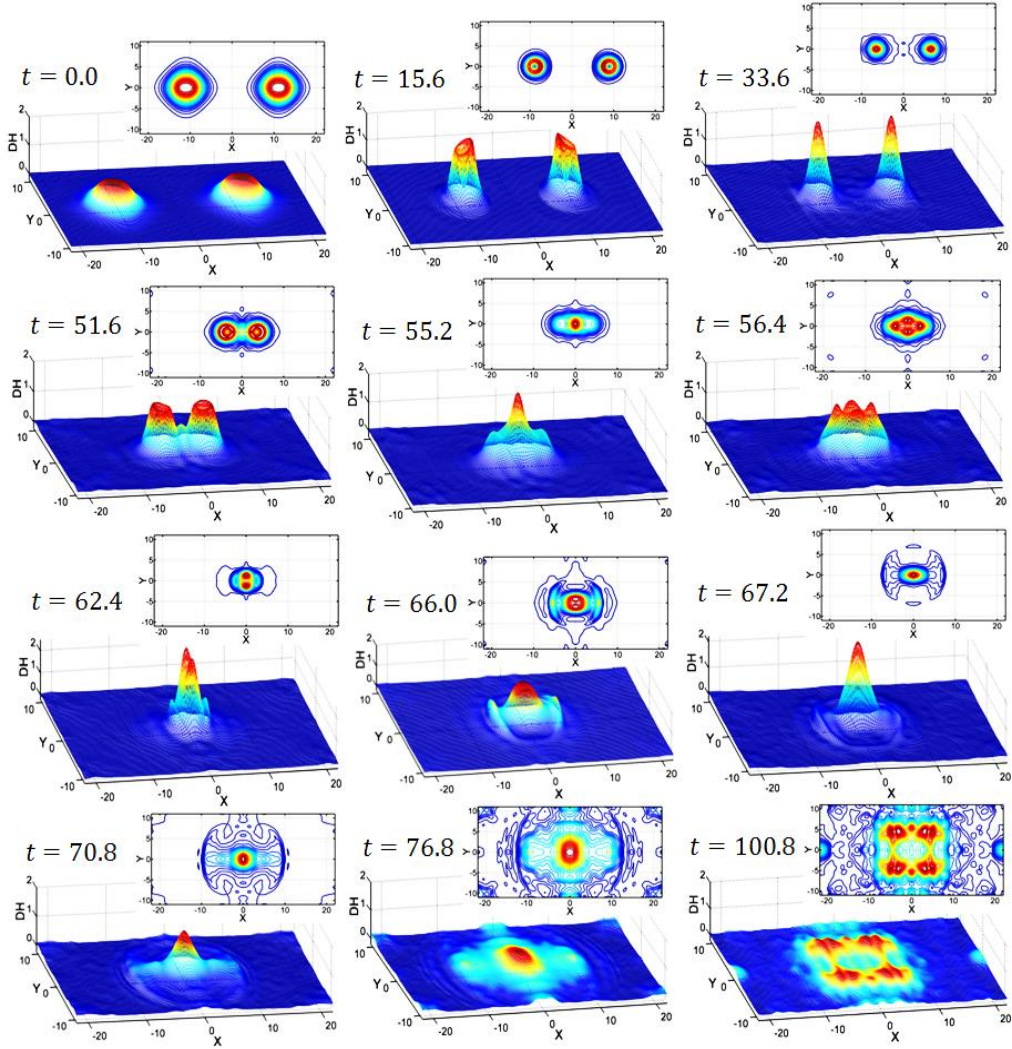

**Fig.5.** Head-on collision of breathers (3) of the SG equation (2), moving at speed $v_{12}(t_0) \approx 0.447$, ($t_0 = 0.0$). Evolution of energy density (DH). Simulation time: $t \in [0.0, 100.8]$.

**2.3.** $v_{12}(t_0) \approx 0.707$ (Fig.6). In this case, unlike the previous experiments, the breathers interaction occurs as expected for solitons of the SG equation, where breathers pass through each other, although at the presence of the radiation of a certain part of energy in the form of radially symmetric linear perturbation waves.

The energy integral of system of interacting breathers preserved with accuracy: $\frac{\Delta En}{En} \approx 10^{-2}$ (Fig.7).

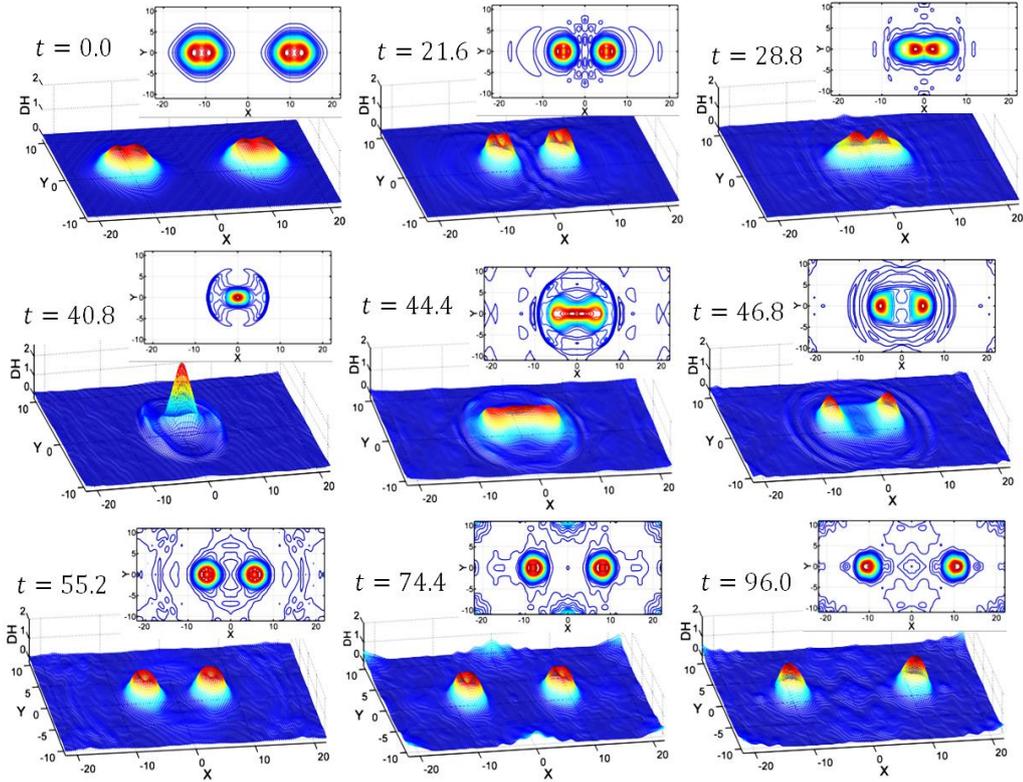

**Fig.6.** Head-on collision of breathers (3) of the SG equation (2), moving at speed $v_{12}(t_0) \approx 0.707$, ($t_0 = 0.0$). Evolution of energy density (DH). Simulation time: $t \in [0.0, 96.0]$.

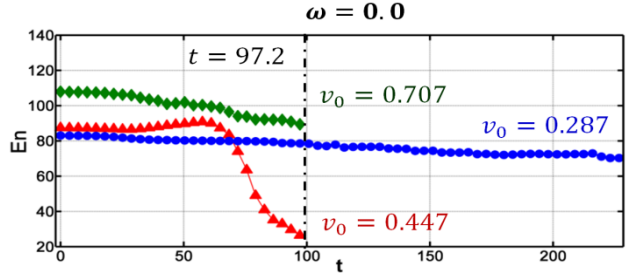

**Fig.7.** The values of energy integral (En) of system of interacting breathers (3) of the SG equation ($\omega = 0.0$), at $v(t_0) \approx 0.287$ (●), $v(t_0) \approx 0.447$ (▲) and $v(t_0) \approx 0.707$ (◆). The total simulation time: $t \in [0.0, 228.0]$.

## III. Interaction of breathers in the O(3) nonlinear sigma model

In this part of the paper present the results of head-on collisions of the breathers of two-dimensional O(3) NSM (1), which were obtained in [7] by introduction to the dynamics of the isotopic spin of fields solutions (3) of the SG equation (2) the specially chosen perturbations:
$$\varphi(x, y, t) = \varphi_0(x, y, t_0) - \omega\tau,$$
$$\omega \neq 0.0.$$

In the following parts of present paper, examples are given for values $\omega = 0.5$ ($\varphi_0 = 0.0$).

**3.1.** $v_{12}(t_0) \approx 0.287$ (Fig.8). In this case, similar to the first experiment of the previous part (see, Fig.4) is observed association of interacting breathers in a single oscillating soliton. The process of interaction in this case in relation to the case described in Fig.4 occurs with a relatively lower energy radiation: $\frac{\Delta En}{En} \approx 10^{-3} - 10^{-2}$ (Fig.11). This is because the breathers of the O(3) NSM derived from breathers of SG equation by adding the perturbations to the vector of A3-field, due to the additional energy and chirality in isospin dynamics are more stable at interactions [7].

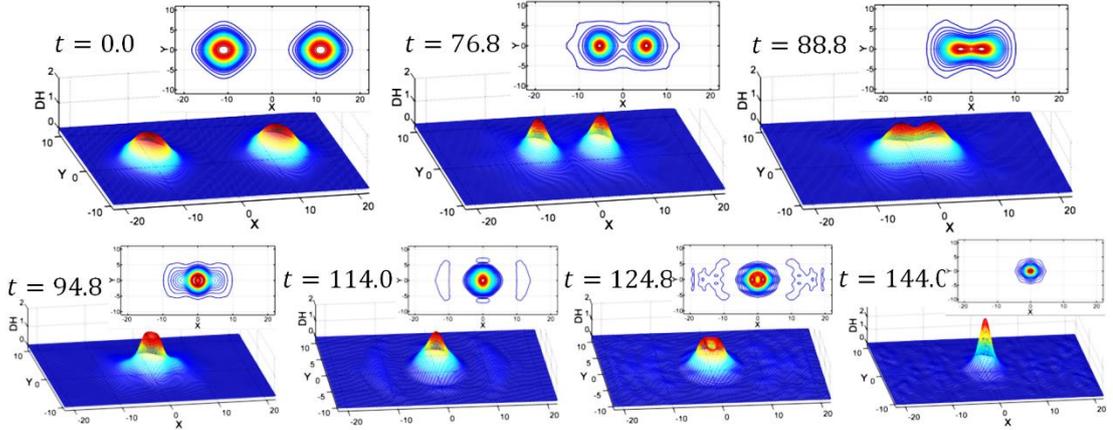

**Fig.8.** Head-on collision of breathers (3) of the O(3) NSM (1), moving at speed $v_{12}(t_0) \approx 0.287$, ($t_0 = 0.0$). Evolution of energy density (DH). Simulation time: $t \in [0.0, 144.0]$.

The following subparts describe similar experiments where the initial speed of motion of the interacting breathers has been increased.

**3.2.** $v_{12}(t_0) \approx 0.447$ (Fig.9). In this case, as in the previous experiment is observed the association of the interacting breathers in a single oscillating soliton. The excess energy is emitted in the form of linear perturbation waves and is absorbed on the edges of the modeling area by special boundary conditions. Note that, in a similar experiment conducted for breathers of the SG equation (see, Fig5, in subpart 3.2), the formed oscillating soliton eventually decay onto the linear waves of perturbations. Thus, it is an additional proof of the relative stability of breathers type of (3) of the O(3) NSM (1) in respect to similar solutions of the SG equation (2).

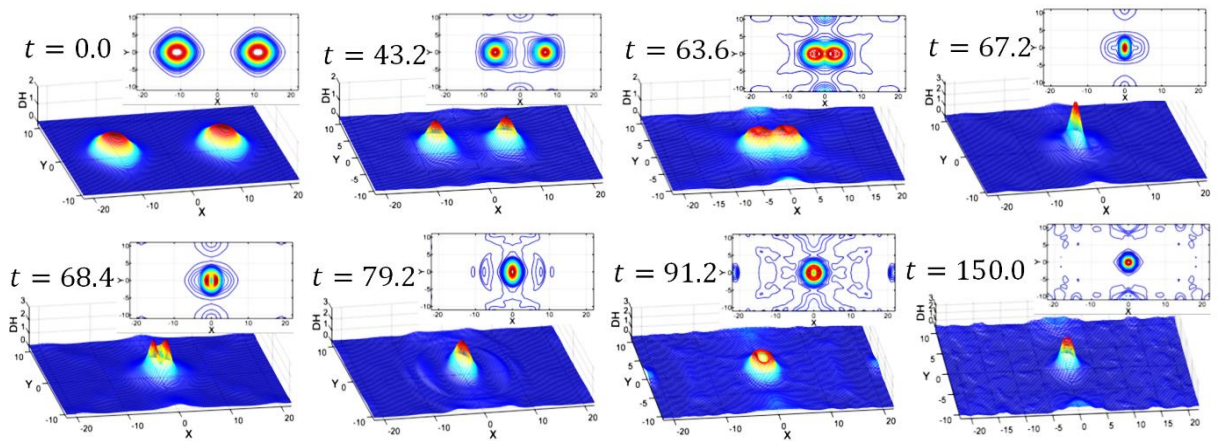

**Fig.9.** Head-on collision of breathers (3) of the O(3) NSM (1), moving at speed $v_{12}(t_0) \approx 0.447$, ($t_0 = 0.0$). Evolution of energy density (DH). Simulation time: $t \in [0.0, 150.0]$.

**3.3.** $v_{12}(t_0) \approx 0.707$ (Fig.10). In this case, similar to the previous two experiments is observed the association of the interacting breathers in a single oscillating soliton. Recall that at similar speeds of breathers movement in the case of the SG equation interacting breathers passed through each other (see, Fig.6). The energy integral of system of interacting breathers preserved with accuracy: $\frac{\Delta En}{En} \approx 10^{-2}$ (Fig.11).

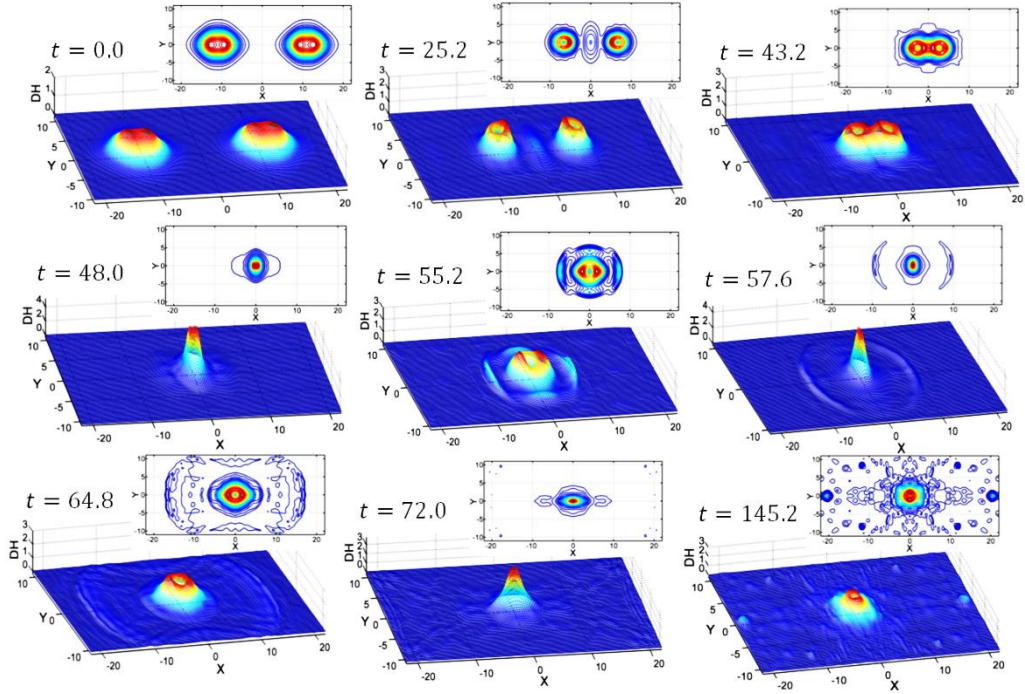

**Fig.10.** Head-on collision of breathers (3) of the O(3) NSM (1), moving at speed $v_{12}(t_0) \approx 0.707$, ($t_0 = 0.0$). Evolution of energy density (DH). Simulation time: $t \in [0.0, 145.2]$.

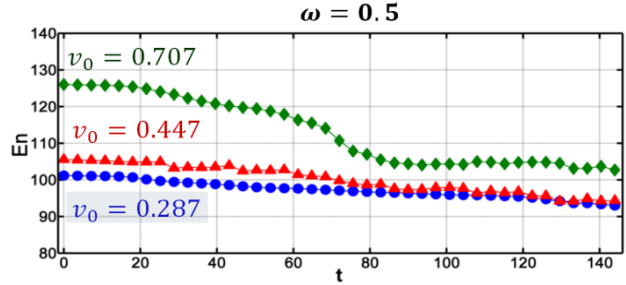

**Fig.11.** The values of the energy integral (En) of system of interacting breathers (3) of the O(3) NSM ($\omega = 0.5$), at $v(t_0) \approx 0.287$ (●), $v(t_0) \approx 0.447$ (▲) and $v(t_0) \approx 0.707$ (♦). The total simulation time: $t \in [0.0, 150.0]$.

## IV. Interaction of breathers in the O(3) nonlinear sigma model at $\omega_1 = -\omega_2$

In this part of the paper considered the model of head-on collisions of breathers (3) of the O(3) NSM (1), with different direction of rotation of the A3-field vector: $\omega_1 = -\omega_2$, at $|\omega| = 0.5$.

**4.1.** $v_{12}(t_0) \approx 0.287$ (Fig.12). In this case, the process of the breathers interaction different from all previous experiments – the breathers at the collision ($100 < t < 120$) reflected from each other ($120 < t < 140$) and continue to move in the opposite direction ($140 < t < 175$).

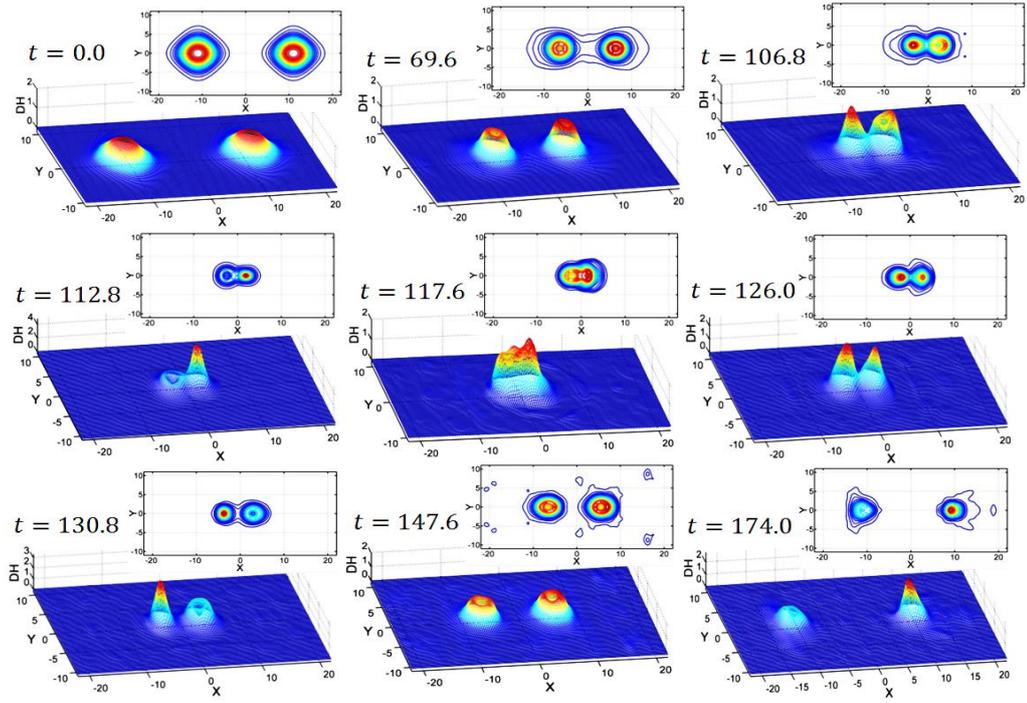

**Fig.12.** Head-on collision of breathers (3) of the O(3) NSM (1), with different values of $\omega$ at $t = 0.0$: $\omega_1 = 0.5$ (in left), $\omega_2 = -0.5$ (in right), moving at speed $v_{12}(t_0) \approx 0.287$, ($t_0 = 0.0$). Evolution of energy density (DH). Simulation time: $t \in [0.0, 174.0]$.

In all experiments of this part the energy integral of system of interacting breathers preserved with accuracy: $\frac{\Delta En}{En} \approx 10^{-3}$ (Fig.15).

Note that in this case there is a minimal loss of energy of interacting breathers (see., e.g., contour projections of energy density (DH) in Fig.12).

**4.2.** $v_{12}(t_0) \approx 0.447$ (Fig.13). In this case, at the collision the breathers pass through each other, but one of the breathers (moving to right, having a value $\omega = 0.5$) by emitting a significant portion of its energy is destroyed ($80 < t < 100$). The second breather after the collision preserved the stable and continues to move in the original direction ($90 < t < 150$).

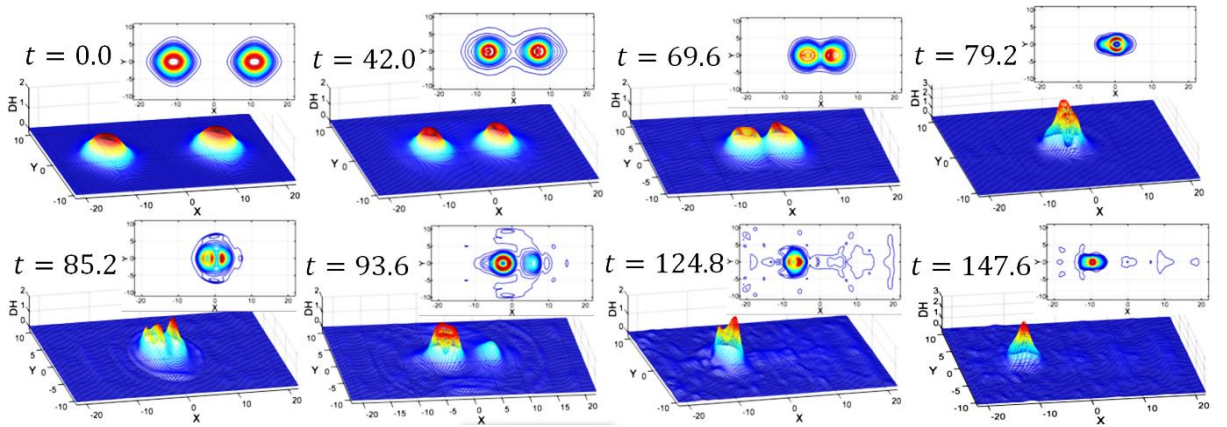

**Fig.13.** Head-on collision of breathers (3) of the O(3) NSM (1), with different values of $\omega$ at $t = 0.0$: $\omega_1 = 0.5$ (in left), $\omega_2 = -0.5$ (in right), moving at speed $v_{12}(t_0) \approx 0.447$, ($t_0 = 0.0$). Evolution of energy density (DH). Simulation time: $t \in [0.0, 147.6]$.

**4.3.** $v_{12}(t_0) \approx 0.707$ (Fig.14). In this experiment, as in the previous case, observed the destruction of one of the interacting breathers (moving to right, with a value $\omega = 0.5$), while $t \in (70.0, 96.0)$. Unlike the previous case, in the collision occurs the resets a large part of the energy ($t = 58.8$), and the attenuation of the disintegrated breather is relatively slow. The second breather after collision, losing a substantial portion of the energy, however, remains stable and continues to move in the original direction ($60 < t < 96$).

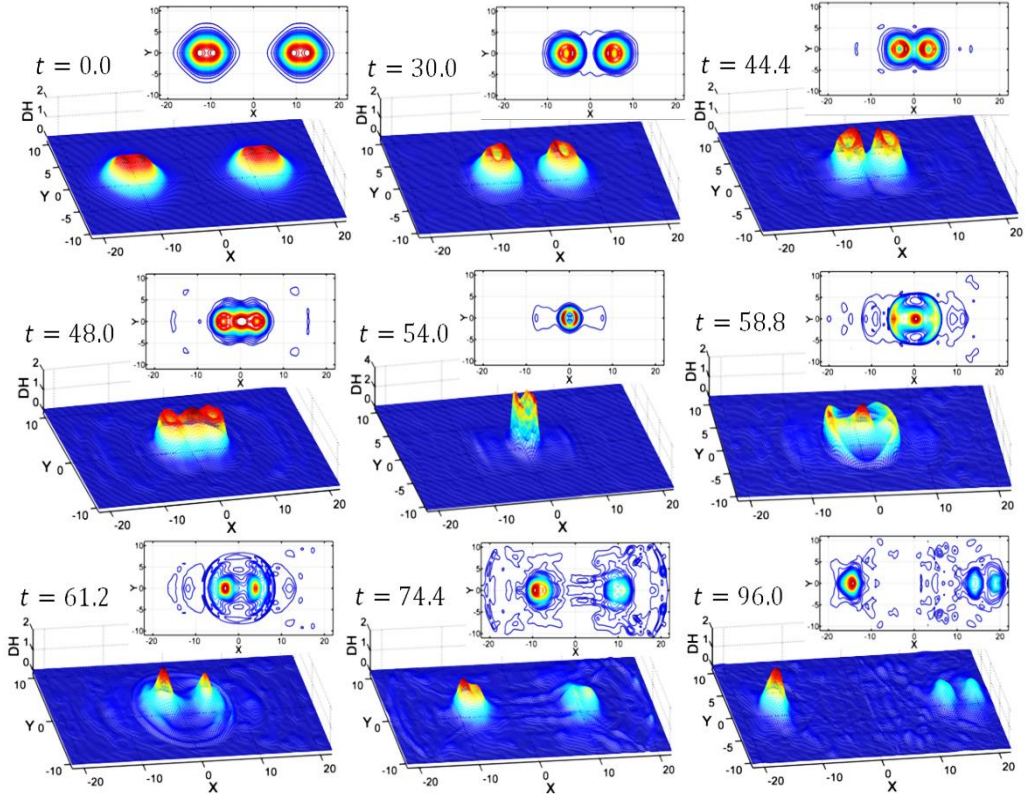

**Fig.14.** Head-on collision of breathers (3) of the O(3) NSM (1), with different values of $\omega$ at $t = 0.0$: $\omega_1 = 0.5$ (in left), $\omega_2 = -0.5$ (in right), moving at speed $v_{12}(t_0) \approx 0.707$, ($t_0 = 0.0$). Evolution of energy density (DH). Simulation time: $t \in [0.0, 96.0]$.

The energy integral of system of interacting breathers preserved with accuracy: $\frac{\Delta En}{En} \approx 10^{-2}$ (Fig.15).

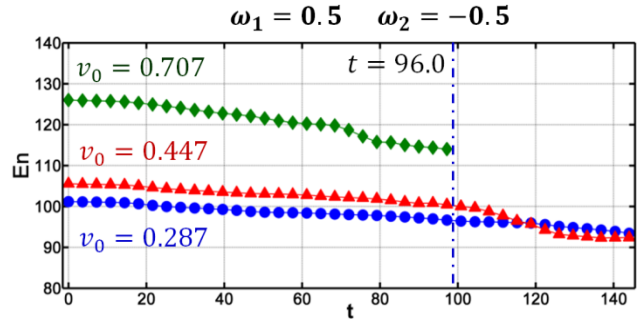

**Fig.15.** The values of the energy integral (En) of system of interacting breathers (3) of the O(3) NSM ($\omega_1 = 0.5, \omega_2 = -0.5$), at $v(t_0) \approx 0.287$ (●), $v(t_0) \approx 0.447$ (▲) and $v(t_0) \approx 0.707$ (◆). The total simulation time: $t \in [0.0, 174.0]$.

## V. Interaction of breathers of the O(3) nonlinear sigma model and sine-Gordon equation

In this part of the paper considered the processes of a frontal collision of breathers of the O(3) NSM (moving in the positive direction of the $x$-axis) and the SG equation (moving in the negative direction of the $x$-axis ) having the values of rotation frequency of A3-field vector $\omega_1 = 0.5$ and $\omega_2 = 0.0$ respectively.

**5.1.** $v_{12}(t_0) \approx 0.287$ (Fig.16). On Fig.16 the breather O(3) NSM at $t = 0.0$ is on the left and the breather of SG equation on right side, and both are moving in the opposite directions. When approaching the resonance zone the breather of the SG equation gradually fades and "absorbed" by the breather field of the O(3) NSM. At this the united breather periodically emits the radially symmetric linear perturbation waves, which are absorbed by the boundary conditions. At $t = 0.0$ the value of the energy integral of system of interacting breathers is equal to $En(t_0) = En_{NSM}(t_0) + En_{SG}(t_0) \approx 92.09965$, where $En_{NSM}(t_0) \approx 50.58921$ and $En_{SG}(t_0) \approx 41.51044$ the values of the energy of breathers of the O(3) NSM and SG equation respectively. The united breather the preserving stability ($En(t_{166.8}) \approx 85.24578$) and evolved until the end of the simulation time: $t = 166.8$.

Thus, in this experiment, two different types of interacting breathers of the O(3) NSM and SG equation, formed a united oscillating soliton. Excess energy is about $\approx 7,5\%$ of the total energy breathers, which emitted in the form of linear radially symmetric waves of perturbations. This union of two interacting breathers occurs in a special way – a breather of the SG equation gradually fading and "absorbed" by the breather field of the O(3) NSM. Next considered the results of similar processes at increased speeds of interacting breathers.

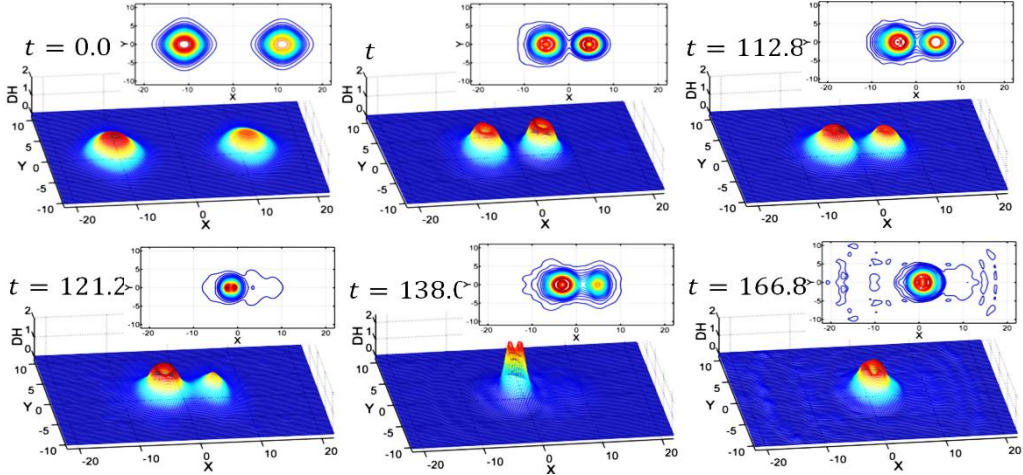

**Fig.16.** Head-on collision of breathers (3) of the O(3) NSM ($\omega_1 = 0.5$, in left, at $t = 0.0$) and SG equation ($\omega_2 = 0.0$, in right, at $t = 0.0$), moving at speed $v_{12}(t_0) \approx 0.287$, ($t_0 = 0.0$). Evolution of energy density (DH). Simulation time: $t \in [0.0, 166.8]$.

**5.2.** $v_{12}(t_0) \approx 0.447$ (Fig.17). The process of interaction of different types of breathers in this case is similar to the previous experiment – in a collision of two breathers formed a single oscillating soliton. The breather of the SG equation "absorbed" by the breather field of the O(3) NSM, which at this emits energy in the form of linear radially symmetric perturbation waves. After the unification of breathers ($t \geq 80.0$) the formed united oscillating soliton radiate the part of its energy and remains stable until the end of the simulation time: $t = 151.2$.

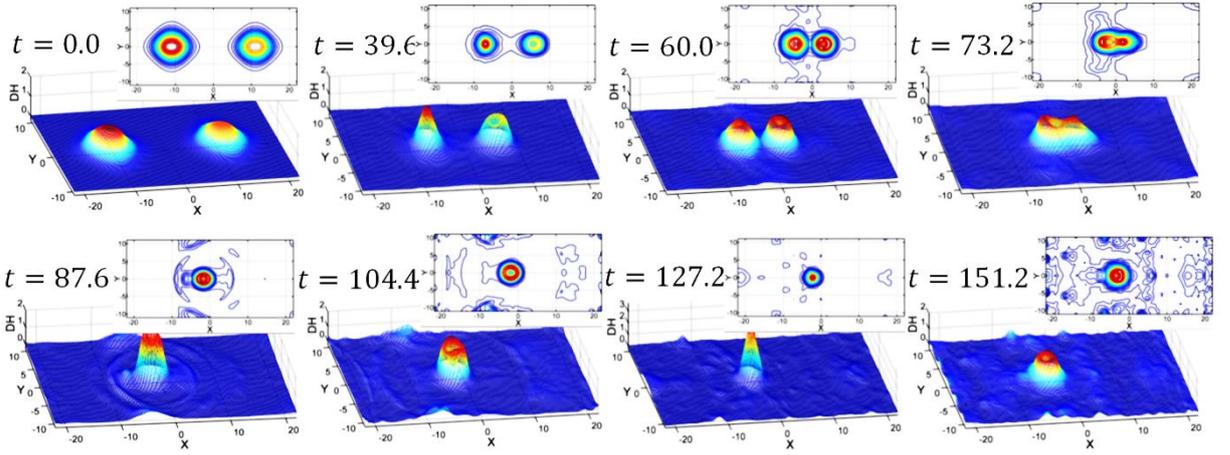

**Fig.17.** Head-on collision of breathers (3) of the O(3) NSM ($\omega_1 = 0.5$, in left, at $t = 0.0$) and SG equation ($\omega_2 = 0.0$, in right, at $t = 0.0$), moving at speed $v_{12}(t_0) \approx 0.447$, ($t_0 = 0.0$). Evolution of energy density (DH). Simulation time: $t \in [0.0, 151.2]$.

**5.3.** $v_{12}(t_0) \approx 0.707$ (Fig.18). In this case, is observed the destruction of one of the interacting breathers after the collision, which continuing to move in a positive direction and rapidly decays ($60 < t < 70$). The second breather remains stable after separation ($t \approx 60.0$) until the end of the simulation time: $t = 122.4$.

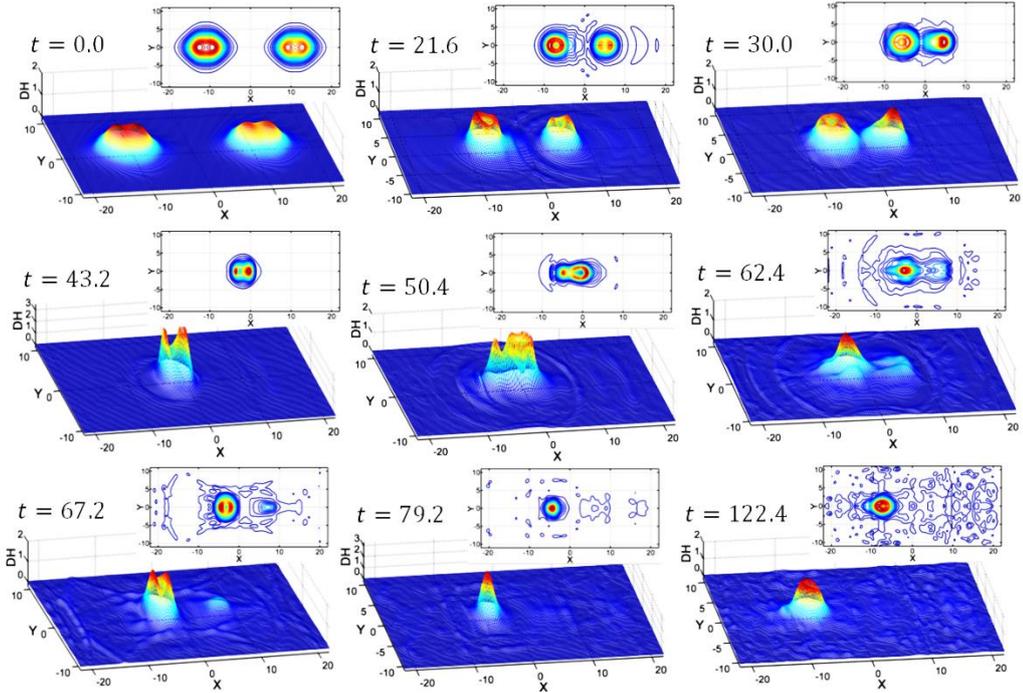

**Fig.18.** Head-on collision of breathers (3) of the O(3) NSM ($\omega_1 = 0.5$, in left, at $t = 0.0$) and SG equation ($\omega_2 = 0.0$, in right, at $t = 0.0$), moving at speed $v_{12}(t_0) \approx 0.707$, ($t_0 = 0.0$). Evolution of energy density (DH). Simulation time: $t \in [0.0, 122.4]$.

In all experiments of this part the energy integral of system of interacting breathers preserved with accuracy: $\frac{\Delta En}{En} \approx 10^{-2}$ (Fig.19).

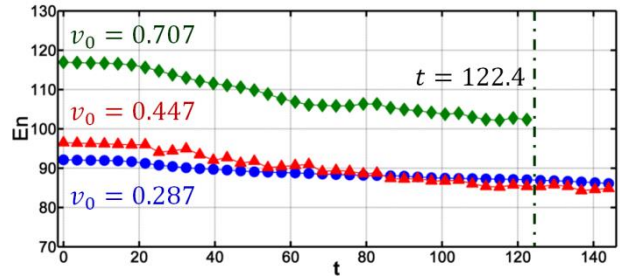

**Fig. 19.** The values of the energy integral (En) of system of interacting breathers (3) of the O(3) NSM ($\omega_1 = 0.5$) and SG equation ($\omega_2 = 0.0$), at $v(t_0) \approx 0.287$ (●), $v(t_0) \approx 0.447$ (▲) and $v(t_0) \approx 0.707$ (◆). The total simulation time: $t \in [0.0, 166.8]$.

## VI. Conclusion

The numerical results presented in this paper are based on the test functions obtained in [7]. This work was motivated by the work [11], where on the basis of Hamiltonian approach were obtained standing and moving two-dimensional breather solutions of the SG equation. On the basis of research methods developed in work [11] in the Lagrangian approach, in [7] were received models of standing and moving two-dimensional oscillating solitons of the O(3) NSM.

In [7] the stability of the obtained solutions at different values of the initial velocity $v_{t_0}$ of their movement and frequency $\omega_{t_0}$ of rotation of isotopic spin vector $S(s_1, s_2, s_3)$ in $S^2$ is shown. These solutions, in [7], were provisionally named breather, due to a certain similarity of their properties with classical breather solutions. The obtained in [7] solutions (3) are periodic; possess internal degrees of freedom and the dynamics sufficiently stable (loss of energy for more than 45000 cycles of iteration is not more than 5.15%).

In the results of numerical simulation of collision of breathers (3) of the SG equation (2), which are presented in the second part of this paper, were obtained the models of inelastic interactions. In our previous works (see., e.g., [5,9]), where carried out similar experiments for the one-dimensional case, at the frontal collision of breathers of the SG equation, solitons pass through each other with little or no loss of energy. Also in cited papers have also been received and interaction models of breather solutions of the two-dimensional O(3) NSM, where from the colliding breathers formed one oscillating soliton. A similar process we are witnessing in the third part of present paper, where a frontal collision breathers (3) of the model (1) are combined into a localized oscillating perturbation (oscillating soliton).

In subpart 4.1 of the fourth part of present paper we received a unique model where at the frontal interaction, occurs collision and mutual reflection of breathers (3). Similar processes (collision and reflection of solitons) were observed in the study of the interaction of two-dimensional topological solitons (see. e.g., [10]). But in the case of dynamic solitons (breathers), this process occurs in our experiments for the first time.

Interesting models have also been received in the last, the fifth part of the present work, at the study of the interaction of different types of breathers of the – O(3) NSM and SG equation. In all experiments of this part, at reducing the distance between the solitons observed the process of attraction and destruction by the breather field of the O(3) NSM the breather of SG equation.

The study of the dynamics of interaction of localized solutions allow manifest fully to their special particle-like properties [12]. Due to the large volume of numerical material in this paper presented only the technical description of the results of experiments that which need to the strict theoretical description. One of the methods to identify the nature of the processes described in present work is the study of the isospin dynamics of a system of interacting soliton fields.

## Acknowledgments

The author is very grateful to Prof. Kh. Kh. Muminov for useful discussions.